\documentclass[aps,prb,reprint,superscriptaddress]{revtex4-2}

\usepackage{amsmath, newtxtext, newtxmath}

\usepackage{braket}
\usepackage{graphicx}
\usepackage{dcolumn}
\usepackage{bm}
\usepackage{soul}
\usepackage{color}
\definecolor{bblue}{rgb}{0, 0.0, 0.8}
\definecolor{rred}{rgb}{0.7, 0.0, 0.0}

\usepackage{hyperref}
\hypersetup{
    colorlinks=true,        
    linkcolor=rred,          
    citecolor=rred,        
    filecolor=rred,      
    urlcolor=rred           
}

\newcommand{\rd}{\textcolor{black}}

\begin{document}

\title{Magnetic field induced valence change in Eu(Co$_{1-x}$Ni$_{x}$)$_{2}$P$_{2}$ up to 60 T}

\author{Raito~Nakamura}\email[Present address: Magnetic Materials Research Center, Shin-Etsu Chemical Co., Ltd., Echizen-shi, Fukui 915-8515, Japan: ]{raito.nkmr@shinetsu.jp}
\author{Azumi~Ishita}
\author{Jin~Nakamura}
\affiliation{Department of Engineering Science, University of Electro-Communications, Chofu, Tokyo 182-8585, Japan}
\author{Hiroto~Ohta}
\affiliation{Department of Molecular Chemistry and Biochemistry, Doshisha University, Kyo-Tanabe, Kyoto 610-0321, Japan}
\author{Yuya~Haraguchi}
\author{Hiroko~Aruga~Katori}
\affiliation{Department of Applied Physics and Chemical Engineering, Tokyo University of Agriculture and Technology, Koganei, Tokyo 184-8588, Japan}
\author{Hajime~Ishikawa}
\author{Akira~Matsuo}
\author{Koichi~Kindo}
\affiliation{Institute for Solid State Physics, University of Tokyo, Kashiwa, Chiba, Japan}
\author{Minoru~Nohara}
\affiliation{Department of Quantum Matter, AdSE, Hiroshima University, Higashi-Hiroshima 739-8530, Japan}
\author{Akihiko~Ikeda}\email[Corresponding author: ]{a-ikeda@uec.ac.jp}
\affiliation{Department of Engineering Science, University of Electro-Communications, Chofu, Tokyo 182-8585, Japan}

\date{\today}

\begin{abstract}
The solid solution 122 compounds, Eu(Co$_{1-x}$Ni$_{x}$)$_{2}$P$_{2}$, show valence transition between divalent state and intermediate valence states at Eu, which is firmly correlated to multiple degrees of freedom in the solid such as the isostructural transition between the collapsed tetragonal (cT) and uncollapsed tetragonal (ucT) structures, $3d$ magnetism, and the formation of P-P dimers.
To gain insights into the correlated behavior, we investigate the effect of high magnetic fields on the samples of $x = 0.4$ and $0.5$ using magnetostriction and magnetization measurements up to 60 T.
The samples are in the Eu valence fluctuating regime, where the possible structural transition from cT to ucT may be induced by the Eu valence change under the magnetic fields.
For both samples, magnetostriction smoothly increases with increasing magnetic fields.
The behavior is in good agreement with the calculated results using the interconfigurational fluctuation (ICF) model that describes the Eu valence change.
This indicates that $\Delta L$ represents the change of the Eu valence state in these compounds.
Magnetization curves for both compounds show good agreement with the ICF model at high magnetic fields.
In contrast, in the low magnetic field region, 
magnetization curves do not agree with the ICF model.
These results indicate that the Eu valence changes manifest themselves in the magnetization curves at high magnetic fields and that the magnetism of the $3d$ electrons manifests itself in the magnetization at low magnetic fields.
Hence, we conclude that the valence change occurs within the Eu valence fluctuation regime coupled with the cT structure.
Thereby, we believe that the transition to ucT structure which is firmly coupled with the divalent Eu state does not occur within the magnetic field range of the present study.
Even higher magnetic fields or pulsed magnetic fields with slower pulse durations may induce such large state changes in the present material, which is triggered by the Eu valence change under high magnetic fields.
\end{abstract}

\maketitle


\section{Introduction}

The 122 compounds with the ThCr$_{2}$Si$_{2}$-type structure are the fertile playground hosting a variety of physical phenomena, ranging from the heavy fermion physics in $4f$ electron systems, high $T_{\rm c}$ superconductivity in Fe pnictides, and itinerant ferromagnets originating in $3d$ electrons of transition metals.
In systems of {\it Ln}{\it Tm}$_{2}$P$_{2}$ ($Ln$: Lanthanoids, {\it Tm}: Transition metals), the characteristic isostructural transition occurs between collapsed tetragonal (cT) and uncollapsed tetragonal structure (ucT).
The transition accompanies forming and breaking of the molecular bond in P dimers.
The P dimer dissociates into 2P$^{3-}$ by filling the anti-bonding orbital ($\sigma^{*}_{\rm P-P}$) with electron-doping in the ucT structure.
The P dimer forms as (P$^{2-}$)$_{2}$ when $\sigma^{*}_{\rm P-P}$ becomes vacant in the cT structure.
The Fermi level of the system is controlled by the $d$ electron number of {\it Tm}
 \cite{HoffmannJPC2002}.
The isostructural in 122 compounds change usually occurs as a continuous transition.
It is a unique feature of Eu{\it Tm}$_{2}$P$_{2}$ systems that the isostructural transition occurs as the first order transition by external stimuli such as temperature \cite{HuhntPhysicaB1997}, pressure \cite{ChefkiPRL1998}, and also by doping Ni to the Co site \cite{NakamuraJPSJ2022}.
This is presumably due to the additional Eu valence degree of freedom in Eu{\it Tm}$_{2}$P$_{2}$ systems.

Recently, under ultrahigh magnetic fields, not only magnetic but also structural transitions are found in systems such as VO$_{2}$ \cite{YHMatsudaNC2020} and solid oxygen \cite{NomuraPRL2014} where strong spin-lattice couplings are in play.
It is of great interest to investigate high magnetic field effects on the correlated multiple degrees of freedom in 122 compounds.

\begin{figure}
\begin{center}
\includegraphics[width = \linewidth, clip]{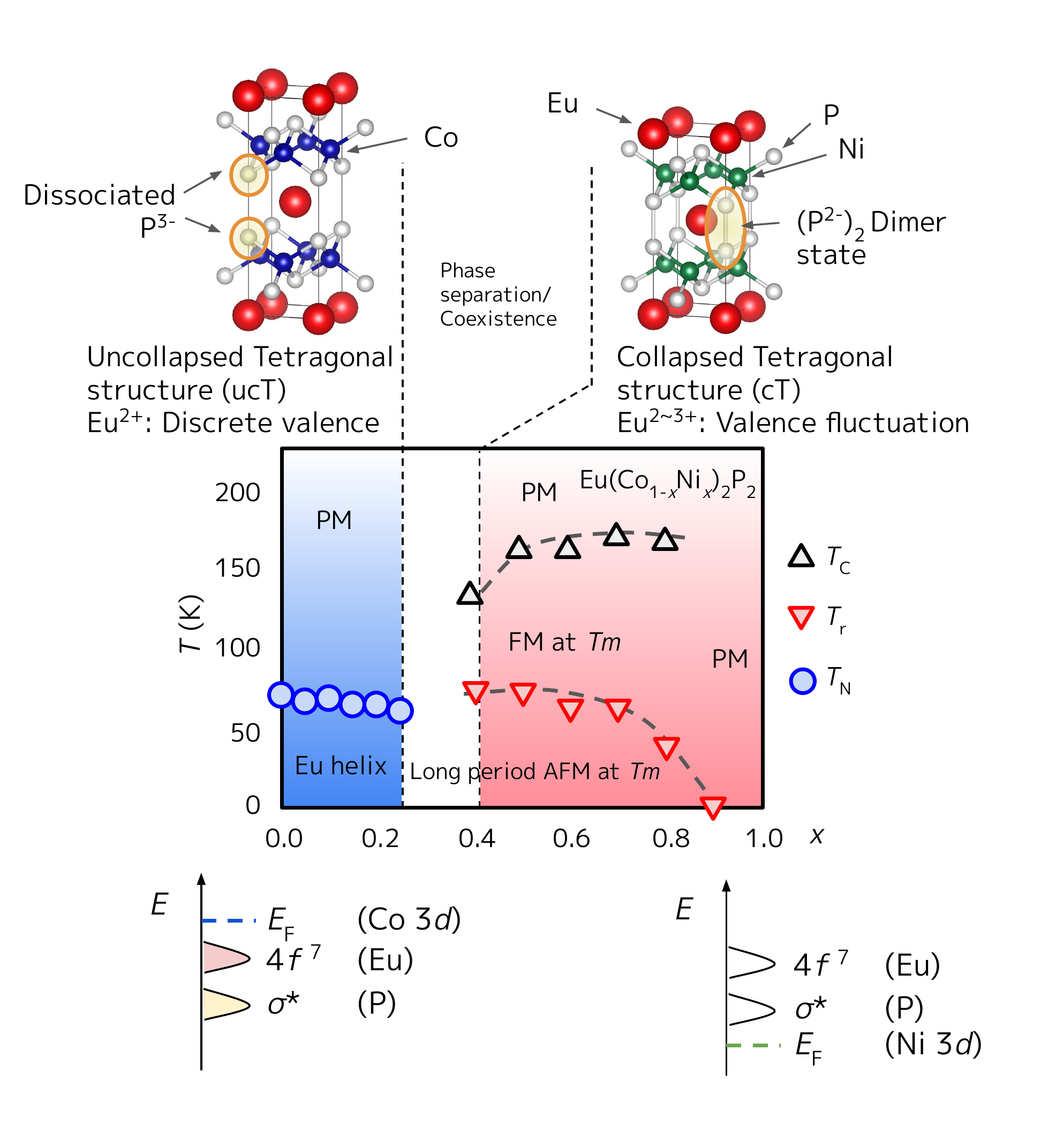} 
\caption{Phase diagram of Eu(Co$_{1-x}$Ni$_{x}$)$_{2}$P$_{2}$ as a function of $x$ based on Refs. \cite{NakamuraJPSJ2022, NakamuraPhDThesis2022}.
$T_{\rm C}$, $T_{\rm r}$, and $T_{\rm N}$ indicate the transition temperatures for the itinerant ferromagnetic order, the reorientation of spins to form the long-period antiferromagnetic order, and the Eu helical magnetic order, respectively.
Schematic energy diagrams for EuCo$_{2}$P$_{2}$ and EuNi$_{2}$P$_{2}$ are shown at the bottom.
The diagrams are based on Ref. \cite{BornickChemMat1994} which assumes the localized limit for $4f^7$ and $\sigma^{*}$ for simplicity.
 \label{intro}}
\end{center}
\end{figure}

Nakamura {\it et al.} investigated the solid solution compounds, Eu(Co$_{1-x}$Ni$_{x}$)$_{2}$P$_{2}$, which show the discontinuous isostructural transitions between cT and ucT structures at $x = 0.25 \sim 0.35$ and various magnetic transitions at low temperatures which are summarised in Fig.~\ref{intro} \cite{NakamuraJPSJ2022, NakamuraPhDThesis2022}.
EuCo$_{2}$P$_{2}$ and EuNi$_{2}$P$_{2}$, take ucT and cT structures, respectively .
The magnetic and electronic properties are quite distinct between these materials.
In EuCo$_{2}$P$_{2}$, the helical spin magnetism arises purely from Eu$^{2+}$.
In contrast, in EuNi$_{2}$P$_{2}$, the itinerant electrons show heavy fermion behavior with the Kondo temperature of 80 K and a strong fluctuation at Eu valence \cite{HiranakaJPSJ2013}.
The Eu valence changes from  Eu$^{2.2+}$ to  Eu$^{2.7+}$ which is the intermediate states between the magnetic Eu$^{2+}$ ($4f^{7}$, $^{8}S_{7/2}$) and nonmagnetic Eu$^{3+}$ ($4f^{6}$, $^{7}F_{0}$).
Several magnetic orders of $3d$ electrons occur in Eu(Co$_{1-x}$Ni$_{x}$)$_{2}$P$_{2}$ at $x=0.4-0.8$ as shown in Fig.~\ref{intro}. 
The electron transfer from Eu($4f$) to {\it Tm}($3d$) states may give rise to the exotic magnetic orderings \cite{NakamuraJPSJ2022}.
Such emergence of magnetism due to the electron transfer from $4f$ state to $3d$ electrons is also reported in the filled-skutterudite compounds Yb$_{x}$Fe$_{4}$Sb$_{12}$ \cite{YamaokaPRL2011}.

In these compounds, such various binary competitions are in play, as the cT-ucT structural instability, the formation and breaking of P dimers, Fermi level tuning through changing the 3$d$ electron numbers with Ni/Co ratio, and the discrete Eu valence and Eu valence fluctuation states.
These factors are strongly coupled with each other.
Especially, the samples of $x = 0.4$ and 0.5 are the typical cases where the magnetism of the $3d$ electrons and Eu valence are coupled \cite{NakamuraJPSJ2022}.

In the present study, we control the valence state of Eu using pulsed high magnetic fields up to 60 T.
Magnetic fields stabilize the Eu$^{2+}$ state over the ground state Eu$^{3+}$ in the samples of Eu(Co$_{1-x}$Ni$_{x}$)$_{2}$P$_{2}$ with $x = 0.4$ and 0.5.
We aim to monitor the valence change by means of magnetostriction measurements under magnetic fields, which is an approach analogues to Ref. \cite{HiranakaJPSJ2013} where the correspondence between the valence change and volume expansion is observed in EuNi$_{2}$P$_{2}$.
The ultimate motivation is to induce the structural transition using high magnetic fields from cT to ucT which is correlated to the Eu valence.
We compare the results of magnetostriction and magnetization to the results of the calculation using the interconfigurational fluctuation (ICF) model.

\section{Methods}
We generate the pulsed magnetic fields up to 60 T using pulse magnets at the MegaGauss Laboratory of the Institute for Solid State Physics, the University of Tokyo, Japan.
We carry out high-resolution magnetostriction measurements using the fiber Bragg grating (FBG) and optical filter method \cite{IkedaRSI2018} and magnetization measurements using the induction method.
In magnetostriction measurements, we use powdered samples.
The powdered samples are solidified using a small amount of epoxy glue, which is then glued to the optical fiber with FBG.
With the two-channel system of our FBG magnetostriction monitor, we can measure the magnetostriction of the two samples of $x = 0.4$ and 0.5 simultaneously in a single pulse.
We measure only the longitudinal magnetostriction.
The present samples are expected to show no magnetic anisotropy, judging from the fact that the magnetic moment of Eu$^{2+}$ arises purely from electron spins, $^{8}S_{7/2}$ ({\it i.e.} no orbital angular momentum), and that Eu$^{3+}$ has no magnetic moment, $^{7}F_{0}$.
\rd{Note that there is a discrepancy between the measured results $\Delta L/L_{\rm{measured}}$ and the intrinsic strain of the sample $\Delta L/L_{\rm{sample}}$, with the relation of $\Delta L/L_{\rm{measured}} = \Delta L/L_{\rm{sample}} \times  p_1 \times  p_2$, where  $0 < p_1, p_2 < 1$.
$p_1$ originates in the path of strain from the sample via glue to the FBG.
$p_2$ originates in that the powder sample is solidified using glue.}

To consider the valence change of Eu as a function of magnetic fields, we analyze the ICF model \cite{MistsudaPRB1997} and compare it to the experimental results.
In the ICF model, the valence fluctuation is effectively taken into account using $T^*$ as $T^* = \sqrt{T^2 + T_{\rm f}^2}$ with $T_{\rm f}$ representing the strength of valence fluctuation.
We define $\beta = (k_{\rm B}T^*)^{-1}$ with $k_{\rm B}$ being Boltzmann factor.
The ground state is Eu$^{3+}$ ($J=0$).
The excited state is Eu$^{2+}$ ($J=7/2$) with the energy gap $E_{\rm ex}$, which are expressed as
\begin{equation}
E_3 = 0 \quad {\rm and} \quad E_2 = g_J \mu_{\rm B} J_{z} B + E_{\rm ex},
\end{equation}
where $g_{J}$, $\mu_{\rm B}$, and $B$, are $g$-factor ($g_J=2$), Bohr magneton, and magnetic fields, respectively.
The partition function is expressed as
\begin{equation}
Z = \exp{\left(-\beta E_3\right)} + \sum_{J_{z}=-7/2}^{7/2} \exp{\left(-\beta E_2\right)}.
\end{equation}
The occupation of Eu$^{2+}$, free energy, and magnetization are deduced as
\begin{align}
P_2 &=  \frac{\sum_{J_{z}} \exp{\left(-\beta E_2\right)}}{Z}, \\
F &= - \beta^{-1} \ln{Z}, \\
M &= - \frac{dF}{dB}.
\end{align}
The valence of Eu is calculated as $v_{\rm Eu} = 3 - P_2$.

\begin{figure}
\begin{center}
\includegraphics[width = \linewidth, clip]{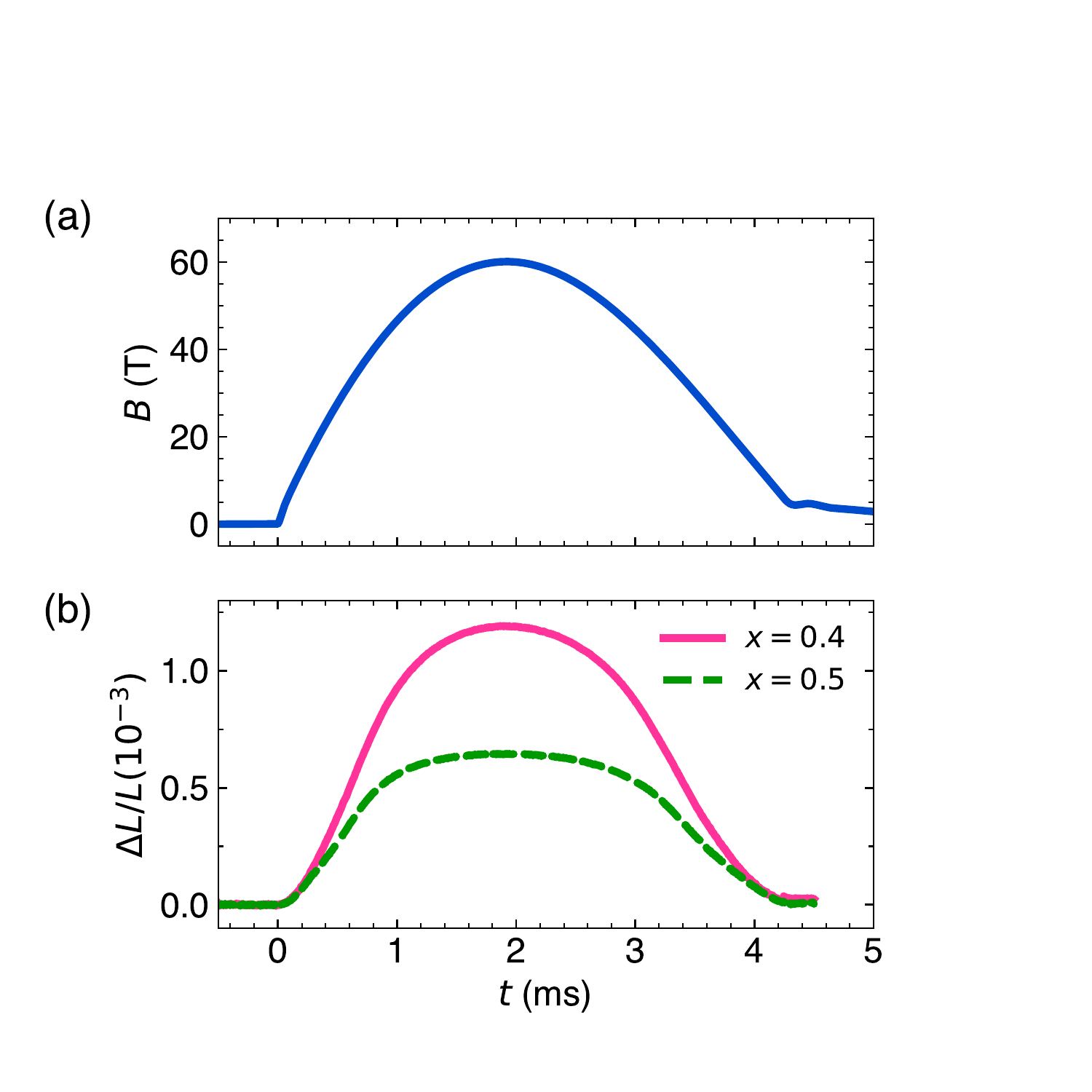} 
\caption{
(a) Pulsed magnetic field and (b) magnetostriction as a function of time for the polycrystalline samples of Eu(Co$_{1-x}$Ni$_{x}$)$_{2}$P$_{2}$ with $x = 0.4$ and 0.5 at the temperature of 4.2 K.
Two magnetostriction curves are measured simultaneously in a single magnetic field pulse utilizing the two-channel FBG magnetostriction measurement system.
 \label{time}}
\end{center}
\end{figure}

\section{Results}

\begin{figure}[h!]
\begin{center}
\includegraphics[width = 0.95\linewidth, clip]{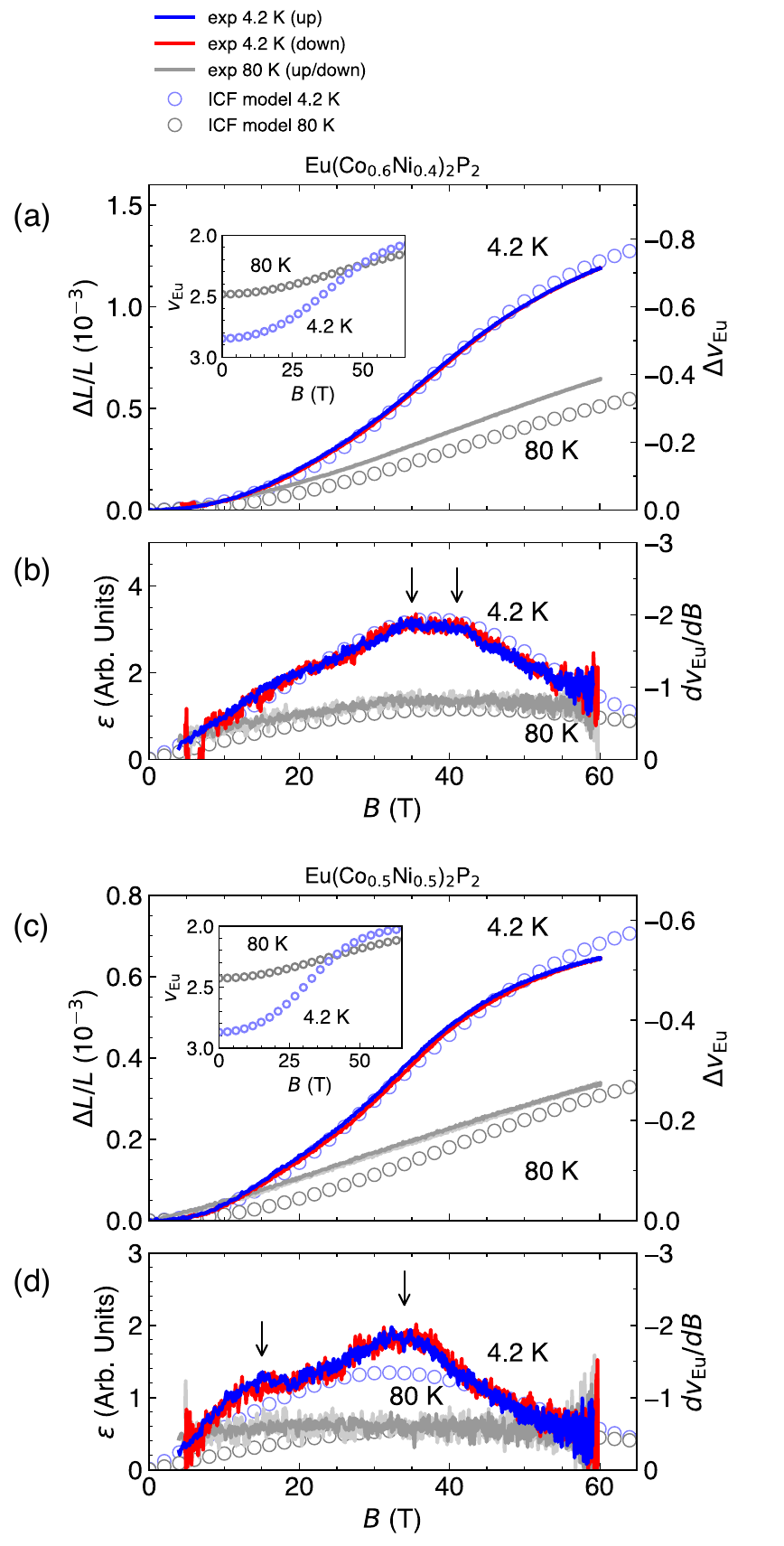} 
\caption{
Magnetostriction ($\Delta L$) and the derivative of magnetostriction $\varepsilon$ $(\equiv d(\Delta L/L) / dB)$ of Eu(Co$_{1-x}$Ni$_{x}$)$_{2}$P$_{2}$ ($x = 0.4$ and $0.5$). (a) $\Delta L$ of $x = 0.4$, (b) $\varepsilon$ of $x = 0.4$, (c) $\Delta L$ of $x = 0.5$, (d) $\varepsilon$ of $x = 0.5$.
Calculated results of Eu valence change $\Delta v_{\rm Eu}$ and its absolute value $v_{\rm Eu}$ are shown in each figure with scales on each right axis, using the ICF model with parameters of ($E_{\rm ex}=190$~K, $T_{\rm f}=50$~K) and ($E_{\rm ex}=160~{\rm K}$, $T_{\rm f}=60~{\rm K}$) for $x=0.4$ and 0.5, respectively.
 \label{L}}
\end{center}
\end{figure}

\begin{figure}[h!]
\begin{center}
\includegraphics[width = 0.9\linewidth, clip]{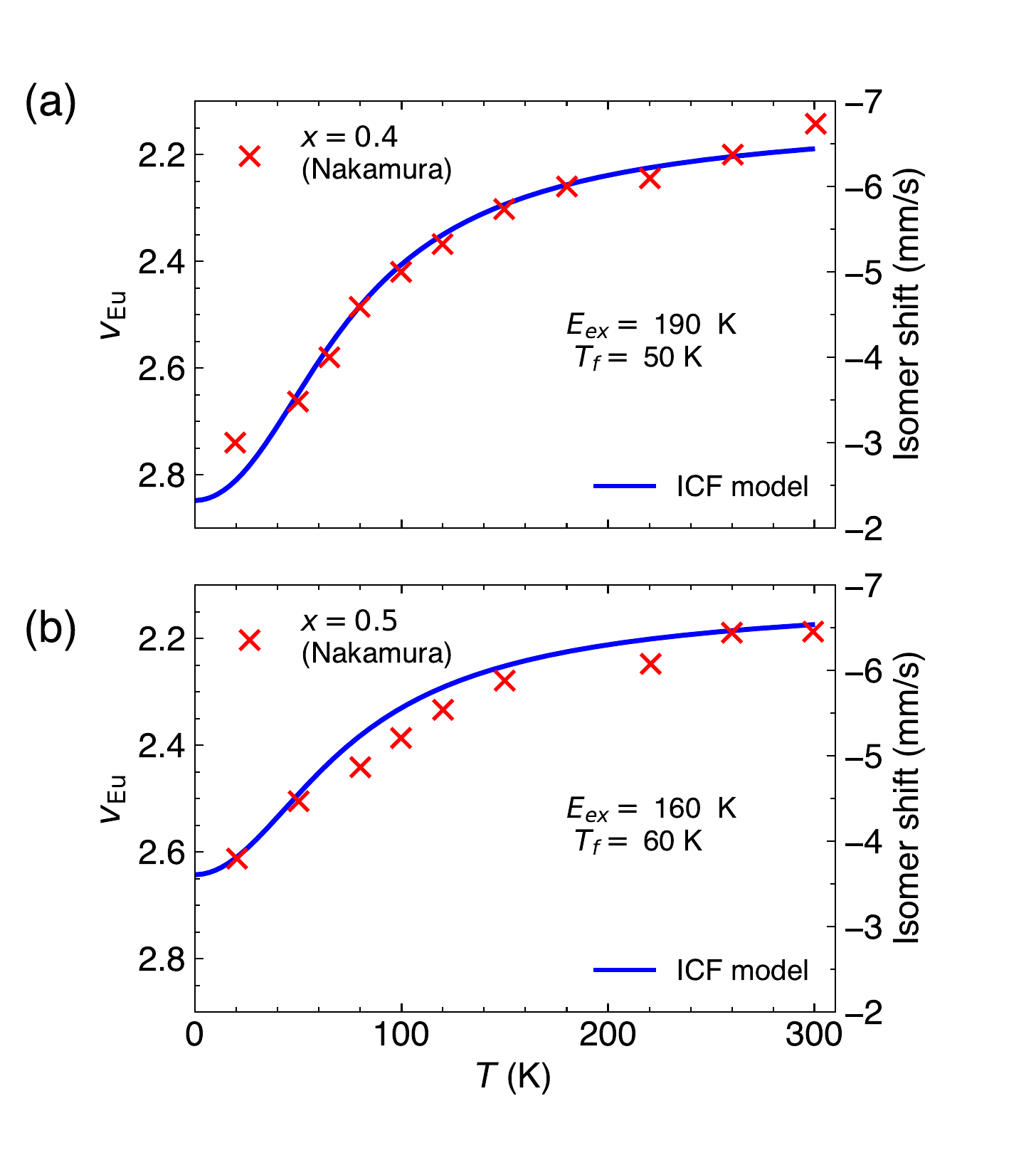} 
\caption{
The solid lines indicate the calculated results of Eu valence $v_{\rm Eu}$ using the ICF model with parameters of ($E_{\rm ex}=190$ K, $T_{\rm f}=50$ K) and ($E_{\rm ex}=160$ K, $T_{\rm f}=60$ K) for (a) $x=0.4$ and (b) $x=0.5$, respectively.
The cross symbols indicate the experimental values of the isomer shift determined using the $^{151}$Eu M\"{o}ssbauer spectroscopy \cite{NakamuraPhDThesis2022}.
\label{mossbauer}}
\end{center}
\end{figure}

\begin{figure}[h!]
\begin{center}
\includegraphics[width = 0.9\linewidth, clip]{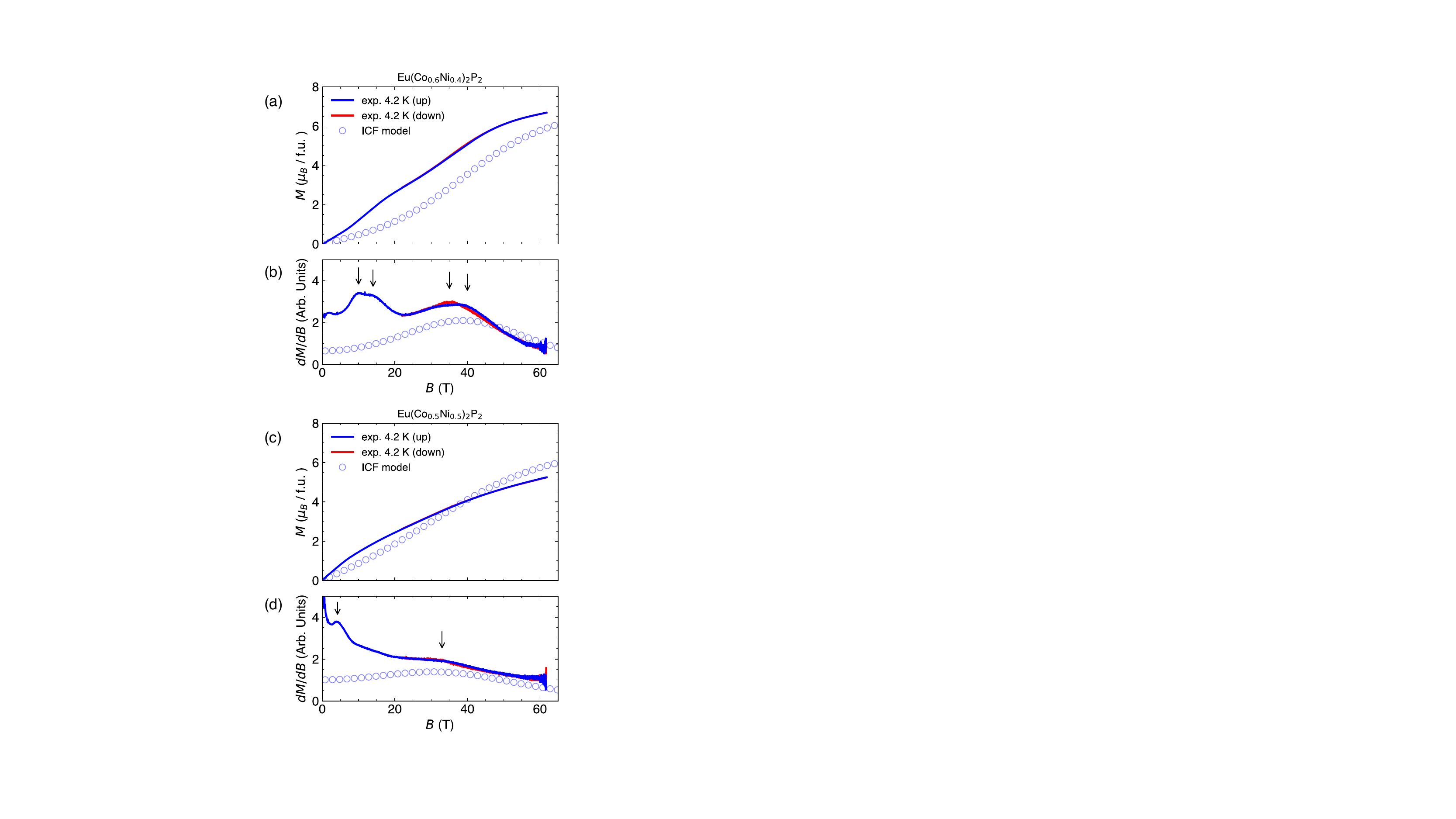} 
\caption{
(a) Magnetization ($M$) and (b) $dM/dB$ of Eu(Co$_{1-x}$Ni$_{x}$)$_{2}$P$_{2}$ at $x = 0.4$, and at (c),(d) $x = 0.5$ are shown with the calculated results of Eu valence change $\Delta v_{\rm Eu}$ and $M$ calculated using the ICF model with parameters of ($E_{\rm ex}=190$ K, $T_{\rm f}=50$ K) and ($E_{\rm ex}=160$ K, $T_{\rm f}=60$ K) for $x=0.4$ and 0.5, respectively.
 \label{MH}}
\end{center}
\end{figure}

Representative data of the pulsed magnetic field and measured magnetostriction ($\Delta L$) is shown in Fig. \ref{time}.
$\Delta L$ curves for the samples of Eu(Co$_{1-x}$Ni$_{x}$)$_{2}$P$_{2}$ with $x = 0.4$ and $x = 0.5$ are measured at the temperature of 4.2 K.
Two kinds of samples are measured simultaneously by making use of the two-channel FBG magnetostriction monitor, which is a minor updated system from Ref. \cite{IkedaRSI2018}.

$\Delta L$ and the derivative of magnetostriction $\varepsilon$ $(\equiv d(\Delta L/L) / dB)$ of the samples of $x = 0.4$ and 0.5 as a function of magnetic fields are shown in Figs.~\ref{L}(a)-\ref{L}(d) with the scales on the left axes.
As one sees in Figs.~\ref{L}(a) and \ref{L}(c), magnetostriction curves continuously increase with increasing magnetic fields.
Figures \ref{L}(b) and \ref{L}(d) show that $\varepsilon$ has a maximum at 39 and 33 T for $x=0.4$ and 0.5, respectively.

The calculated results of the valence change $\Delta v_{\rm Eu} = v_{\rm Eu}(B) - v_{\rm Eu}(B = 0{\rm \ T}) $ and its derivative $d v_{\rm Eu}/dB$ are plotted in the same figures with the scales on the right axes.
The insets in Figs.~\ref{L}(a) and \ref{L}(c) show the absolute value of the Eu valence $v_{\rm Eu}$ as a function of $B$.
The parameters of the ICF model for $x = 0.4$ and 0.5 are ($E_{\rm ex} = 190$ K, $T_{\rm f} = 50$ K), and ($E_{\rm ex} = 160$ K, $T_{\rm f} = 60$ K), respectively.
\rd{The calculated valence changes are fitted to the experimental result of $\Delta L/L$ by using the shape of the curves.}

As shown in Figs.~\ref{L}(a)-\ref{L}(d), one notices that the calculated results using the ICF model are in good agreement with the experimental results.
This indicates that the change of $\Delta L$ with increasing magnetic fields well represents the change of Eu valence, considering the fact that the ICF model represents the Eu valence change.

To verify the obtained parameters in the calculations, we compare the valence change as a function of temperatures in the ICF model with the reported isomer shift of the $^{151}$Eu M\"{o}ssbauer spectra as shown in Figs.~\ref{mossbauer}(a) and \ref{mossbauer}(b).
With the parameter sets for $x=0.4$ and 0.5, we calculate the temperature dependence of the Eu valence state using the ICF model.
The experimental results of the isomer shift of the $^{151}$Eu M\"{o}ssbauer spectroscopy represent the microscopic change of Eu valence in the solids.
For both samples of $x=0.4$ and 0.5, they qualitatively agree with the experimental results.
The agreement of the ICF model to the isomer shift of the $^{151}$Eu M\"{o}ssbauer spectra supports our notion that $\Delta L$ well represents well the Eu valence state under high magnetic fields.

Figures \ref{MH}(a)-\ref{MH}(d) show that magnetization and $dM/dB$ as a function of magnetic fields for the samples of $x=0.4$ and 0.5 are shown with the scales on the left axes.
In the figures, the calculated results of $M$ using the ICF model with the same parameter sets introduced earlier are shown with the scales on the right axes.

In the data with the sample of $x=0.4$, one can notice multiple peaks in $dM/dB$ data which are indicated by the arrows in Figs.~\ref{MH}(a) and \ref{MH}(b).
The peak structures in $dM/dB$ at 35 and 39 T show qualitative agreement with the peak structure in the calculated $dM/dB$, indicating the peak structure originates from the change of Eu valence.
By contrast, the peak structure in $dM/dB$ at 11 and 15 T are not present in the calculated $dM/dB$.
This indicates that the peak structures in $dM/dB$ at 11 and 15 T are assumed to arise from another origin.
Nakamura {\it et al.} reported this feature arises from the spin-flip transitions in $3d$ electrons from the Co/Ni site \cite{NakamuraJPSJ2022}.

The overall features are similar in $x=0.5$ data to those in $x=0.4$ data as shown in Figs.~\ref{MH}(c) and \ref{MH}(d).
The experimental curves of $M$ and $dM/dB$ agree well with the calculated results using the ICF model at high magnetic fields.
By contrast, they do not fit well in the low magnetic field region.
This indicates the magnetism of $3d$ electrons of transition metals is in play at the low magnetic field region.
The emergence of peculiar transition from a ferromagnetic to an antiferromagnetic phase with decreasing temperatures is reported earlier \cite{NakamuraJPSJ2022}, which is due to the $3d$ electron states of the transition metal sites, which is a characteristic behavior in the cT structures in this system.
We also note that the magnetism of $3d$ electrons shows little influence on magnetostriction, indicating a small spin-lattice coupling of $3d$ electrons in the present systems.

We observe the small hysteresis in the magnetization process as shown in Figs.~\ref{MH}(a) and \ref{MH}(b) with the peak in the up-sweep of the pulsed magnetic field at $B=39$ T and that in the down sweep at $B=35$ T, which infers a weak first-order transition.
On the other hand, we do not observe the corresponding hysteresis in the magnetostriction data as shown in Figs.~\ref{L}(a) and \ref{L}(b).
We speculate that this discrepancy is due to the higher resolution of magnetization measurement than magnetostriction measurement to such a subtle feature in the present experimental condition.

\section{Discussion}

Here, we discuss the valence change and its coupling to the structural change.
$\Delta v_{\rm Eu}$ yields $-0.75$ at 60 T as shown in Figs.~\ref{L}(a) and \ref{L}(c).
The inset shows that this $\Delta v_{\rm Eu}$ corresponds to the change from 2.85 to 2.10.
This is a similar amount of Eu valence change with the temperature dependence of $v_{\rm Eu}$ as shown in Fig.~\ref{mossbauer}(a).
Besides, $M$ grows up to 6.6 $\mu_{\rm B}$ / f.u. at 60 T which is very close to the value of the saturation magnetization of isolated Eu$^{2+}$, 7.0 $\mu_{\rm B}$ / f.u.
These results indicate that the magnetic field of 60 T changes the Eu valence very close to the $2.0+$.
The data for $x=0.4$ and $x=0.5$ are qualitatively similar.
At 60 T, $\Delta L/L$ shows the value of $1.2 \times 10^{-3}$ and $0.65 \times 10^{-3}$ for $x=0.4$ and $x=0.5$, respectively.
The absolute value of $\Delta L/L$ in the present study may be smaller than the intrinsic $\Delta L/L$ of the sample due to the fact that we measured the powdered sample solidified with epoxy glue.
Thus, we can not directly compare the observed amount of $\Delta L/L$ in the order of $ \sim 10^{-3}$ with the expected value of $\sim6 \%$ which is the reported change of lattice volume \cite{NakamuraJPSJ2022, HuhntPRB1997, ChefkiPRL1998}.

Based on the agreement of the data to the ICF model, we conclude that the presently observed lattice changes originated in the valence change of Eu within the valence fluctuating regime whose lattice state is in the cT region.
Thereby, no isostructural transition from cT to ucT is induced by the magnetic fields in the present study.
More likely, the presently observed transition may have the same root as the weak metamagnetic transition observed in EuNi$_2$P$_2$ at around 30 T \cite{HiranakaJPSJ2013}.
This is also supported by the absence of the change of lattice in the strong first-order manner expected for the cT - ucT transition.
The first-order change of lattice in the cT - ucT transition is reported in the solid solution study of the present system \cite{NakamuraJPSJ2022} and the pressure-induced transition \cite{HuhntPRB1997, ChefkiPRL1998}.

The absence of the structural transition from cT to ucT contrasts the fact that the valence change yields very close to 2.0 and that the Eu$^{2+}$ state is coupled to the ucT structure.
Our speculation is that with even higher magnetic fields that induce the complete Eu$^{2+}$ state, one should observe the isostructural transition from cT to ucT by magnetic fields.
There are two possible origins of the required higher magnetic fields.
One is that the internal energy of the ucT phase is still higher, so we need higher a magnetic field.
\rd{In this case, one may need $B> \sim 450$ T estimated with the relation of Zeeman energy and the internal energy difference, $B\Delta M > \Delta E_{\rm{cT-ucT}}$.
We tentatively estimated $\Delta M \sim 1 \mu_{\rm{B}}$/f.u. and $\Delta E_{\rm{cT-ucT}} > 300$ K based on the facts that $M$ is far from the saturation for $\sim1 \mu_{\rm{B}}$/f.u. and that the cT-ucT transition is not induced in the temperature range up to 300 K.
}
Another possibility is that the isostructural transition involves a large magnetic field hysteresis region, which is indicated by the coexistence region of $x=0.25 - 0.35$ in Fig.~\ref{intro}.
In the latter case, not only the higher magnetic fields, but also the slower pulse sweep rate, higher temperature, and the stimulation by laser pulse may induce the first-order phase transition.
A study at higher magnetic fields is possible by means of a 1000 T environment generated using electro-magnetic flux compression technique \cite{NakamuraRSI2018} and the high-speed magnetostriction monitor \cite{IkedaRSI2017, IkedaPRL2020, IkedaNC2023}.
Further, a microscopic measurement on lattice is also possible at 100 T range by using the portable destructive pulse magnet and x-ray free electron laser facilities \cite{IkedaAPL2022}.
Such studies are underway.

The obtained value of $E_{\rm ex} = 160$ K for the sample of $x=0.5$ is smaller than the obtained value of $E_{\rm ex} = 190$ K for the sample of $x=0.4$.
This is qualitatively in good agreement with the fact that position of the peaks in $\varepsilon$ and $dM/dB$ for the sample of $x=0.5$ appear at a lower magnetic field than that for the sample of $x=0.4$ as shown in Figs. \ref{L}(b), \ref{L}(d), \ref{MH}(b), and \ref{MH}(d).
This observation may indicate that the valence and magnetization saturate at a lower magnetic field with $x=0.5$ than in the case of $x=0.4$.
In reality, the valence of the sample with $x=0.5$ starts to saturate at lower magnetic fields than the sample with $x=0.4$ as shown by the $\Delta L$ curves in Figs. \ref{L}(a) and \ref{L}(c).
In contrast, $M$ of the sample with $x=0.5$ is still far from the saturation value of 7 $\mu_{\rm B}$ / f.u. even at 60 T, while the $M$ of the sample with $x=0.4$ show saturation behavior already at 60 T as shown in Figs. \ref{MH}(a) and \ref{MH}(c).
Such reduction of magnetization for the sample with $x=0.5$ at high magnetic fields may be a sign of the strong Kondo coupling ($c$-$f$ hybridization) which is significant at $x = 1.0$ \cite{HiranakaJPSJ2013}.
On the other hand, the sample with $x=0.4$ show almost fully saturated magnetization.
It infers that the Kondo coupling is not significant at $x=0.4$.

Under magnetic fields, the electron transfer to Eu occurs resulting in the valence change from the 3+ side to the 2+ side.
We note that this electron transfer does not occur directly among Eu and the P dimers.
Both the Eu$^{2+}$ states and the dissociated 2(P$^{3-}$) states are coupled to the ucT structure, both of which are electron-doped states.
Thus, these electrons are presumably transferred from the transition metal site.
The schematic drawings of energy levels in Fig. \ref{intro} clearly shows that the electron configuration is controlled by the Fermi level determined by the $3d$ electron state.

We compare the present system with the valence fluctuating systems of EuNi$_2$(Si$_{1-x}$Ge$_x$)$_2$, where Eu valence is fluctuating at $x=0.75 - 0.82$ \cite{WadaJPCM1997}.
The magnetic fields induce valence change with first-order transition \cite{YHMatsudaPRL2009}.
The valence change is reproducible using the ICF model but with an additional parameter as $E_{\rm ex} = E_{0}(1-\alpha P_2)$, where $E_{0}$ and $\alpha$ are excited energy without the cooperative effect and strength of the cooperative effect, respectively \cite{WadaJPCM1997}.
The insertion of $P_2$ into $E_{\rm ex}$ represents the cooperative effect \cite{CroftPRL1982} between Eu$^{2+}$ that transforms the crossover into the first order transition.
Such cooperative effects are absent in the present system considering the fact that the field-induced change is a gradual one.

In various systems of Eu{\it Tm}$_2$(Si$_{1-x}$Ge$_x$)$_2$, the valence transitions with variations of temperature and magnetic fields are reported \cite{MistsudaPRB1997, WadaJPCM1997}.
The magnetic fields for the Eu valence transition show linear dependences on the distance between Eu and {\it Tm} \cite{MistsudaPRB1997}.
Similarly, in the systems of Eu{\it Tm}$_2$As$_2$, the isomer shift of the $^{151}$Eu M\"{o}ssbauer spectroscopy shows a linear dependence on the Eu-{\it Tm} distances \cite{RaffiusJPCS1993}, indicating that the Eu-{\it Tm} distances are an important determining factor in these systems.
The present study is the first case showing the magnetic field-induced valence transition in the Eu 122 compounds with Phosphorus.
We note that the Eu-{\it Tm} length is also an important parameter in the systems with phosphorus or arsenic \cite{RaffiusJPCS1991, NakamuraPhDThesis2022}.
This point will become clear by the study of magnetic field-induced valence transition in such systems as EuRh$_{2}$(P$_{x}$As$_{1-x}$)$_{2}$ \cite{BarlaPRB2004}.

\section{Conclusion}
We report the magnetization and magnetostriction measurements of Eu(Co$_{1-x}$Ni$_{x}$)$_{2}$P$_{2}$ with $x = 0.4$ and 0.5.
We compare the results with the calculated results of Eu valence using the ICF model.
The magnetostriction curves are in good agreement with the valence change calculated using the ICF model, indicating that magnetostriction originates in the Eu valence change.
In fact, the obtained parameters of $E_{\rm ex}$ and $T_{\rm f}$ and the ICF model also show good agreement with the reported temperature dependence of the isomer shift of the $^{151}$Eu M\"{o}ssbauer spectroscopy.
The present valence change occurred within the Eu valence fluctuation region coupled to the cT structure, not involving the field-induced cT - ucT isostructural transition, which may require even higher magnetic fields.

\begin{acknowledgements}
AI is supported by MEXT Leading Initiative for Excellent Young Researchers Grant Number JPMXS0320210021, JST FOREST Program (Grant Number JPMJFR222W, Japan), JSPS Grant-in-Aid for Transformative Research Areas Grant Number 23H04861, Grant-in-Aid for Scientific Research (B) 23H01121.
HI is supported by the JSPS KAKENHI Grants No. 22H04467.
The 3D images of crystal structure in Fig.~\ref{intro} are rendered
using VESTA3 \cite{MommaJAC2011}.
\end{acknowledgements}

\bibliography{FBG_Eu122}

\begin{thebibliography}{25}%
\makeatletter
\providecommand \@ifxundefined [1]{%
 \@ifx{#1\undefined}
}%
\providecommand \@ifnum [1]{%
 \ifnum #1\expandafter \@firstoftwo
 \else \expandafter \@secondoftwo
 \fi
}%
\providecommand \@ifx [1]{%
 \ifx #1\expandafter \@firstoftwo
 \else \expandafter \@secondoftwo
 \fi
}%
\providecommand \natexlab [1]{#1}%
\providecommand \enquote  [1]{``#1''}%
\providecommand \bibnamefont  [1]{#1}%
\providecommand \bibfnamefont [1]{#1}%
\providecommand \citenamefont [1]{#1}%
\providecommand \href@noop [0]{\@secondoftwo}%
\providecommand \href [0]{\begingroup \@sanitize@url \@href}%
\providecommand \@href[1]{\@@startlink{#1}\@@href}%
\providecommand \@@href[1]{\endgroup#1\@@endlink}%
\providecommand \@sanitize@url [0]{\catcode `\\12\catcode `\$12\catcode
  `\&12\catcode `\#12\catcode `\^12\catcode `\_12\catcode `\%12\relax}%
\providecommand \@@startlink[1]{}%
\providecommand \@@endlink[0]{}%
\providecommand \url  [0]{\begingroup\@sanitize@url \@url }%
\providecommand \@url [1]{\endgroup\@href {#1}{\urlprefix }}%
\providecommand \urlprefix  [0]{URL }%
\providecommand \Eprint [0]{\href }%
\providecommand \doibase [0]{https://doi.org/}%
\providecommand \selectlanguage [0]{\@gobble}%
\providecommand \bibinfo  [0]{\@secondoftwo}%
\providecommand \bibfield  [0]{\@secondoftwo}%
\providecommand \translation [1]{[#1]}%
\providecommand \BibitemOpen [0]{}%
\providecommand \bibitemStop [0]{}%
\providecommand \bibitemNoStop [0]{.\EOS\space}%
\providecommand \EOS [0]{\spacefactor3000\relax}%
\providecommand \BibitemShut  [1]{\csname bibitem#1\endcsname}%
\let\auto@bib@innerbib\@empty
\bibitem [{\citenamefont {Hoffmann}\ and\ \citenamefont
  {Zheng}(2002)}]{HoffmannJPC2002}%
  \BibitemOpen
  \bibfield  {author} {\bibinfo {author} {\bibfnamefont {R.}~\bibnamefont
  {Hoffmann}}\ and\ \bibinfo {author} {\bibfnamefont {C.}~\bibnamefont
  {Zheng}},\ }\bibfield  {title} {\bibinfo {title} {{\rm Making and Breaking
  Bonds in the Solid State: The ThCr$_{2}$Si$_{2}$ Structure}},\ }\href
  {https://doi.org/10.1021/j100266a007} {\bibfield  {journal} {\bibinfo
  {journal} {J. Phys. Chem.}\ }\textbf {\bibinfo {volume} {89}},\ \bibinfo
  {pages} {4175} (\bibinfo {year} {2002})}\BibitemShut {NoStop}%
\bibitem [{\citenamefont {Huhnt}\ \emph
  {et~al.}(1997{\natexlab{a}})\citenamefont {Huhnt}, \citenamefont {Michels},
  \citenamefont {Roepke}, \citenamefont {Schlabitz}, \citenamefont {Wurth},
  \citenamefont {Johrendt},\ and\ \citenamefont {Mewis}}]{HuhntPhysicaB1997}%
  \BibitemOpen
  \bibfield  {author} {\bibinfo {author} {\bibfnamefont {C.}~\bibnamefont
  {Huhnt}}, \bibinfo {author} {\bibfnamefont {G.}~\bibnamefont {Michels}},
  \bibinfo {author} {\bibfnamefont {M.}~\bibnamefont {Roepke}}, \bibinfo
  {author} {\bibfnamefont {W.}~\bibnamefont {Schlabitz}}, \bibinfo {author}
  {\bibfnamefont {A.}~\bibnamefont {Wurth}}, \bibinfo {author} {\bibfnamefont
  {D.}~\bibnamefont {Johrendt}},\ and\ \bibinfo {author} {\bibfnamefont
  {A.}~\bibnamefont {Mewis}},\ }\bibfield  {title} {\bibinfo {title}
  {\rm{First-order phase transitions in the ThCr$_{2}$Si$_{2}$-type phosphides
  ARh$_{2}$P$_{2}$ (A = Sr, Eu)}},\ }\href
  {https://doi.org/10.1016/s0921-4526(97)00431-6} {\bibfield  {journal}
  {\bibinfo  {journal} {Physica B}\ }\textbf {\bibinfo {volume} {240}},\
  \bibinfo {pages} {26} (\bibinfo {year} {1997}{\natexlab{a}})}\BibitemShut
  {NoStop}%
\bibitem [{\citenamefont {Chefki}\ \emph {et~al.}(1998)\citenamefont {Chefki},
  \citenamefont {Abd-Elmeguid}, \citenamefont {Micklitz}, \citenamefont
  {Huhnt}, \citenamefont {Schlabitz}, \citenamefont {Reehuis},\ and\
  \citenamefont {Jeitschko}}]{ChefkiPRL1998}%
  \BibitemOpen
  \bibfield  {author} {\bibinfo {author} {\bibfnamefont {M.}~\bibnamefont
  {Chefki}}, \bibinfo {author} {\bibfnamefont {M.~M.}\ \bibnamefont
  {Abd-Elmeguid}}, \bibinfo {author} {\bibfnamefont {H.}~\bibnamefont
  {Micklitz}}, \bibinfo {author} {\bibfnamefont {C.}~\bibnamefont {Huhnt}},
  \bibinfo {author} {\bibfnamefont {W.}~\bibnamefont {Schlabitz}}, \bibinfo
  {author} {\bibfnamefont {M.}~\bibnamefont {Reehuis}},\ and\ \bibinfo {author}
  {\bibfnamefont {W.}~\bibnamefont {Jeitschko}},\ }\bibfield  {title} {\bibinfo
  {title} {\rm{Pressure-induced Transition of the Sublattice Magnetization in
  EuCo$_{2}$P$_{2}$: Change from Local Moment Eu ($4f$) to Itinerant Co ($3d$)
  Magnetism}},\ }\href {https://doi.org/10.1103/PhysRevLett.80.802} {\bibfield
  {journal} {\bibinfo  {journal} {Phys. Rev. Lett.}\ }\textbf {\bibinfo
  {volume} {80}},\ \bibinfo {pages} {802} (\bibinfo {year} {1998})}\BibitemShut
  {NoStop}%
\bibitem [{\citenamefont {Nakamura}\ \emph {et~al.}(2022)\citenamefont
  {Nakamura}, \citenamefont {Ohta}, \citenamefont {Kobayashi}, \citenamefont
  {Haraguchi}, \citenamefont {Katori},\ and\ \citenamefont
  {Nakamura}}]{NakamuraJPSJ2022}%
  \BibitemOpen
  \bibfield  {author} {\bibinfo {author} {\bibfnamefont {R.}~\bibnamefont
  {Nakamura}}, \bibinfo {author} {\bibfnamefont {H.}~\bibnamefont {Ohta}},
  \bibinfo {author} {\bibfnamefont {Y.}~\bibnamefont {Kobayashi}}, \bibinfo
  {author} {\bibfnamefont {Y.}~\bibnamefont {Haraguchi}}, \bibinfo {author}
  {\bibfnamefont {H.~A.}\ \bibnamefont {Katori}},\ and\ \bibinfo {author}
  {\bibfnamefont {J.}~\bibnamefont {Nakamura}},\ }\bibfield  {title} {\bibinfo
  {title} {\rm{Magnetic Properties of Eu(Co$_{1-x}$Ni$_{x}$)$_{2}$P$_{2}$}},\
  }\href {https://doi.org/10.7566/jpsj.91.024701} {\bibfield  {journal}
  {\bibinfo  {journal} {J. Phys. Soc. Jpn.}\ }\textbf {\bibinfo {volume}
  {91}},\ \bibinfo {pages} {024701} (\bibinfo {year} {2022})}\BibitemShut
  {NoStop}%
\bibitem [{\citenamefont {Matsuda}\ \emph {et~al.}(2020)\citenamefont
  {Matsuda}, \citenamefont {Nakamura}, \citenamefont {Ikeda}, \citenamefont
  {Takeyama}, \citenamefont {Suga}, \citenamefont {Nakahara},\ and\
  \citenamefont {Muraoka}}]{YHMatsudaNC2020}%
  \BibitemOpen
  \bibfield  {author} {\bibinfo {author} {\bibfnamefont {Y.~H.}\ \bibnamefont
  {Matsuda}}, \bibinfo {author} {\bibfnamefont {D.}~\bibnamefont {Nakamura}},
  \bibinfo {author} {\bibfnamefont {A.}~\bibnamefont {Ikeda}}, \bibinfo
  {author} {\bibfnamefont {S.}~\bibnamefont {Takeyama}}, \bibinfo {author}
  {\bibfnamefont {Y.}~\bibnamefont {Suga}}, \bibinfo {author} {\bibfnamefont
  {H.}~\bibnamefont {Nakahara}},\ and\ \bibinfo {author} {\bibfnamefont
  {Y.}~\bibnamefont {Muraoka}},\ }\bibfield  {title} {\bibinfo {title}
  {\rm{Magnetic-field-induced insulator-metal transition in W-doped VO$_{2}$ at
  500 T}},\ }\href {https://doi.org/10.1038/s41467-020-17416-w} {\bibfield
  {journal} {\bibinfo  {journal} {Nat. Commun.}\ }\textbf {\bibinfo {volume}
  {11}},\ \bibinfo {pages} {3591} (\bibinfo {year} {2020})}\BibitemShut
  {NoStop}%
\bibitem [{\citenamefont {Nomura}\ \emph {et~al.}(2014)\citenamefont {Nomura},
  \citenamefont {Matsuda}, \citenamefont {Takeyama}, \citenamefont {Matsuo},
  \citenamefont {Kindo}, \citenamefont {Her},\ and\ \citenamefont
  {Kobayashi}}]{NomuraPRL2014}%
  \BibitemOpen
  \bibfield  {author} {\bibinfo {author} {\bibfnamefont {T.}~\bibnamefont
  {Nomura}}, \bibinfo {author} {\bibfnamefont {Y.~U.}\ \bibnamefont {Matsuda}},
  \bibinfo {author} {\bibfnamefont {S.}~\bibnamefont {Takeyama}}, \bibinfo
  {author} {\bibfnamefont {A.}~\bibnamefont {Matsuo}}, \bibinfo {author}
  {\bibfnamefont {K.}~\bibnamefont {Kindo}}, \bibinfo {author} {\bibfnamefont
  {J.~U.}\ \bibnamefont {Her}},\ and\ \bibinfo {author} {\bibfnamefont {T.~U.}\
  \bibnamefont {Kobayashi}},\ }\bibfield  {title} {\bibinfo {title} {\rm{Novel
  Phase of Solid Oxygen Induced by Ultrahigh Magnetic Fields}},\ }\href
  {https://doi.org/10.1103/PhysRevLett.112.247201} {\bibfield  {journal}
  {\bibinfo  {journal} {Phys. Rev. Lett.}\ }\textbf {\bibinfo {volume} {112}},\
  \bibinfo {pages} {247201} (\bibinfo {year} {2014})}\BibitemShut {NoStop}%
\bibitem [{\citenamefont {Nakamura}(2022)}]{NakamuraPhDThesis2022}%
  \BibitemOpen
  \bibfield  {author} {\bibinfo {author} {\bibfnamefont {R.}~\bibnamefont
  {Nakamura}},\ }\href@noop {} {\bibinfo {type} {\rm{PhD Thesis, University of
  Electro-Communications}}} (\bibinfo {year} {2022})\BibitemShut {NoStop}%
\bibitem [{\citenamefont {Bornick}\ and\ \citenamefont
  {Stacy}(1994)}]{BornickChemMat1994}%
  \BibitemOpen
  \bibfield  {author} {\bibinfo {author} {\bibfnamefont {R.~M.}\ \bibnamefont
  {Bornick}}\ and\ \bibinfo {author} {\bibfnamefont {A.~M.}\ \bibnamefont
  {Stacy}},\ }\bibfield  {title} {\bibinfo {title} {\rm{Intermediate Valence in
  EuCo$_{2-x}$Ni$_{x}$P$_{2}$: Interdependence of Structure and Energetics}},\
  }\href {https://doi.org/10.1021/cm00039a014} {\bibfield  {journal} {\bibinfo
  {journal} {Chem. Mater.}\ }\textbf {\bibinfo {volume} {6}},\ \bibinfo {pages}
  {333} (\bibinfo {year} {1994})}\BibitemShut {NoStop}%
\bibitem [{\citenamefont {Hiranaka}\ \emph {et~al.}(2013)\citenamefont
  {Hiranaka}, \citenamefont {Nakamura}, \citenamefont {Hedo}, \citenamefont
  {Takeuchi}, \citenamefont {Mori}, \citenamefont {Hirose}, \citenamefont
  {Mitamura}, \citenamefont {Sugiyama}, \citenamefont {Hagiwara}, \citenamefont
  {Nakama},\ and\ \citenamefont {\={O}nuki}}]{HiranakaJPSJ2013}%
  \BibitemOpen
  \bibfield  {author} {\bibinfo {author} {\bibfnamefont {Y.}~\bibnamefont
  {Hiranaka}}, \bibinfo {author} {\bibfnamefont {A.}~\bibnamefont {Nakamura}},
  \bibinfo {author} {\bibfnamefont {M.}~\bibnamefont {Hedo}}, \bibinfo {author}
  {\bibfnamefont {T.}~\bibnamefont {Takeuchi}}, \bibinfo {author}
  {\bibfnamefont {A.}~\bibnamefont {Mori}}, \bibinfo {author} {\bibfnamefont
  {Y.}~\bibnamefont {Hirose}}, \bibinfo {author} {\bibfnamefont
  {K.}~\bibnamefont {Mitamura}}, \bibinfo {author} {\bibfnamefont
  {K.}~\bibnamefont {Sugiyama}}, \bibinfo {author} {\bibfnamefont
  {M.}~\bibnamefont {Hagiwara}}, \bibinfo {author} {\bibfnamefont
  {T.}~\bibnamefont {Nakama}},\ and\ \bibinfo {author} {\bibfnamefont
  {Y.}~\bibnamefont {\={O}nuki}},\ }\bibfield  {title} {\bibinfo {title}
  {\rm{Heavy Fermion State Based on the Kondo Effect in EuNi$_{2}$P$_{2}$}},\
  }\href {https://doi.org/10.7566/jpsj.82.083708} {\bibfield  {journal}
  {\bibinfo  {journal} {J. Phys. Soc. Jpn.}\ }\textbf {\bibinfo {volume}
  {82}},\ \bibinfo {pages} {083708} (\bibinfo {year} {2013})}\BibitemShut
  {NoStop}%
\bibitem [{\citenamefont {Yamaoka}\ \emph {et~al.}(2011)\citenamefont
  {Yamaoka}, \citenamefont {Jarrige}, \citenamefont {Tsujii}, \citenamefont
  {Lin}, \citenamefont {Ikeno}, \citenamefont {Isikawa}, \citenamefont
  {Nishimura}, \citenamefont {Higashinaka}, \citenamefont {Sato}, \citenamefont
  {Hiraoka}, \citenamefont {Ishii},\ and\ \citenamefont
  {Tsuei}}]{YamaokaPRL2011}%
  \BibitemOpen
  \bibfield  {author} {\bibinfo {author} {\bibfnamefont {H.}~\bibnamefont
  {Yamaoka}}, \bibinfo {author} {\bibfnamefont {I.}~\bibnamefont {Jarrige}},
  \bibinfo {author} {\bibfnamefont {N.}~\bibnamefont {Tsujii}}, \bibinfo
  {author} {\bibfnamefont {J.~F.}\ \bibnamefont {Lin}}, \bibinfo {author}
  {\bibfnamefont {T.}~\bibnamefont {Ikeno}}, \bibinfo {author} {\bibfnamefont
  {Y.}~\bibnamefont {Isikawa}}, \bibinfo {author} {\bibfnamefont
  {K.}~\bibnamefont {Nishimura}}, \bibinfo {author} {\bibfnamefont
  {R.}~\bibnamefont {Higashinaka}}, \bibinfo {author} {\bibfnamefont
  {H.}~\bibnamefont {Sato}}, \bibinfo {author} {\bibfnamefont {N.}~\bibnamefont
  {Hiraoka}}, \bibinfo {author} {\bibfnamefont {H.}~\bibnamefont {Ishii}},\
  and\ \bibinfo {author} {\bibfnamefont {K.~D.}\ \bibnamefont {Tsuei}},\
  }\bibfield  {title} {\bibinfo {title} {\rm{Strong coupling between $4f$
  valence instability and $3d$ ferromagnetism in Yb$_{x}$Fe$_4$Sb$_{12}$
  studied by resonant x-ray emission spectroscopy}},\ }\href
  {https://doi.org/10.1103/PhysRevLett.107.177203} {\bibfield  {journal}
  {\bibinfo  {journal} {Phys. Rev. Lett.}\ }\textbf {\bibinfo {volume} {107}},\
  \bibinfo {pages} {177203} (\bibinfo {year} {2011})}\BibitemShut {NoStop}%
\bibitem [{\citenamefont {Ikeda}\ \emph {et~al.}(2018)\citenamefont {Ikeda},
  \citenamefont {Matsuda},\ and\ \citenamefont {Tsuda}}]{IkedaRSI2018}%
  \BibitemOpen
  \bibfield  {author} {\bibinfo {author} {\bibfnamefont {A.}~\bibnamefont
  {Ikeda}}, \bibinfo {author} {\bibfnamefont {Y.~H.}\ \bibnamefont {Matsuda}},\
  and\ \bibinfo {author} {\bibfnamefont {H.}~\bibnamefont {Tsuda}},\ }\bibfield
   {title} {\bibinfo {title} {\rm{Note: Optical filter method for
  high-resolution magnetostriction measurement using fiber Bragg grating under
  millisecond-pulsed high magnetic fields at cryogenic temperatures}},\ }\href
  {https://doi.org/10.1063/1.5034035} {\bibfield  {journal} {\bibinfo
  {journal} {Rev. Sci. Instrum.}\ }\textbf {\bibinfo {volume} {89}},\ \bibinfo
  {pages} {096103} (\bibinfo {year} {2018})}\BibitemShut {NoStop}%
\bibitem [{\citenamefont {Mitsuda}\ \emph {et~al.}(1997)\citenamefont
  {Mitsuda}, \citenamefont {Wada}, \citenamefont {Shiga}, \citenamefont
  {Aruga~Katori},\ and\ \citenamefont {Goto}}]{MistsudaPRB1997}%
  \BibitemOpen
  \bibfield  {author} {\bibinfo {author} {\bibfnamefont {A.}~\bibnamefont
  {Mitsuda}}, \bibinfo {author} {\bibfnamefont {H.}~\bibnamefont {Wada}},
  \bibinfo {author} {\bibfnamefont {M.}~\bibnamefont {Shiga}}, \bibinfo
  {author} {\bibfnamefont {H.}~\bibnamefont {Aruga~Katori}},\ and\ \bibinfo
  {author} {\bibfnamefont {T.}~\bibnamefont {Goto}},\ }\bibfield  {title}
  {\bibinfo {title} {\rm{Field-induced valence transition of
  Eu(Pd$_{1-x}$Pt$_{x}$)$_{2}$Si$_{2}$}},\ }\href
  {https://doi.org/10.1103/PhysRevB.55.12474} {\bibfield  {journal} {\bibinfo
  {journal} {Phys. Rev. B}\ }\textbf {\bibinfo {volume} {55}},\ \bibinfo
  {pages} {12474} (\bibinfo {year} {1997})}\BibitemShut {NoStop}%
\bibitem [{\citenamefont {Huhnt}\ \emph
  {et~al.}(1997{\natexlab{b}})\citenamefont {Huhnt}, \citenamefont {Schlabitz},
  \citenamefont {Wurth}, \citenamefont {Mewis},\ and\ \citenamefont
  {Reehuis}}]{HuhntPRB1997}%
  \BibitemOpen
  \bibfield  {author} {\bibinfo {author} {\bibfnamefont {C.}~\bibnamefont
  {Huhnt}}, \bibinfo {author} {\bibfnamefont {W.}~\bibnamefont {Schlabitz}},
  \bibinfo {author} {\bibfnamefont {A.}~\bibnamefont {Wurth}}, \bibinfo
  {author} {\bibfnamefont {A.}~\bibnamefont {Mewis}},\ and\ \bibinfo {author}
  {\bibfnamefont {M.}~\bibnamefont {Reehuis}},\ }\bibfield  {title} {\bibinfo
  {title} {\rm{First-order phase transitions in EuCo$_{2}$P$_{2}$ and
  SrNi$_{2}$P$_{2}$}},\ }\href {https://doi.org/10.1103/PhysRevB.56.13796}
  {\bibfield  {journal} {\bibinfo  {journal} {Phys. Rev. B}\ }\textbf {\bibinfo
  {volume} {56}},\ \bibinfo {pages} {13796} (\bibinfo {year}
  {1997}{\natexlab{b}})}\BibitemShut {NoStop}%
\bibitem [{\citenamefont {Nakamura}\ \emph {et~al.}(2018)\citenamefont
  {Nakamura}, \citenamefont {Ikeda}, \citenamefont {Sawabe}, \citenamefont
  {Matsuda},\ and\ \citenamefont {Takeyama}}]{NakamuraRSI2018}%
  \BibitemOpen
  \bibfield  {author} {\bibinfo {author} {\bibfnamefont {D.}~\bibnamefont
  {Nakamura}}, \bibinfo {author} {\bibfnamefont {A.}~\bibnamefont {Ikeda}},
  \bibinfo {author} {\bibfnamefont {H.}~\bibnamefont {Sawabe}}, \bibinfo
  {author} {\bibfnamefont {Y.~H.}\ \bibnamefont {Matsuda}},\ and\ \bibinfo
  {author} {\bibfnamefont {S.}~\bibnamefont {Takeyama}},\ }\bibfield  {title}
  {\bibinfo {title} {\rm{Record indoor magnetic field of 1200 T generated by
  electromagnetic flux-compression}},\ }\href
  {https://doi.org/10.1063/1.5044557} {\bibfield  {journal} {\bibinfo
  {journal} {Rev. Sci. Instrum.}\ }\textbf {\bibinfo {volume} {89}},\ \bibinfo
  {pages} {095106} (\bibinfo {year} {2018})}\BibitemShut {NoStop}%
\bibitem [{\citenamefont {Ikeda}\ \emph {et~al.}(2017)\citenamefont {Ikeda},
  \citenamefont {Nomura}, \citenamefont {Matsuda}, \citenamefont {Tani},
  \citenamefont {Kobayashi}, \citenamefont {Watanabe},\ and\ \citenamefont
  {Sato}}]{IkedaRSI2017}%
  \BibitemOpen
  \bibfield  {author} {\bibinfo {author} {\bibfnamefont {A.}~\bibnamefont
  {Ikeda}}, \bibinfo {author} {\bibfnamefont {T.}~\bibnamefont {Nomura}},
  \bibinfo {author} {\bibfnamefont {Y.~H.}\ \bibnamefont {Matsuda}}, \bibinfo
  {author} {\bibfnamefont {S.}~\bibnamefont {Tani}}, \bibinfo {author}
  {\bibfnamefont {Y.}~\bibnamefont {Kobayashi}}, \bibinfo {author}
  {\bibfnamefont {H.}~\bibnamefont {Watanabe}},\ and\ \bibinfo {author}
  {\bibfnamefont {K.}~\bibnamefont {Sato}},\ }\bibfield  {title} {\bibinfo
  {title} {\rm{High-speed 100 MHz strain monitor using fiber Bragg grating and
  optical filter for magnetostriction measurements under ultrahigh magnetic
  fields}},\ }\href {https://doi.org/10.1063/1.4999452} {\bibfield  {journal}
  {\bibinfo  {journal} {Rev. Sci. Instrum.}\ }\textbf {\bibinfo {volume}
  {88}},\ \bibinfo {pages} {083906} (\bibinfo {year} {2017})}\BibitemShut
  {NoStop}%
\bibitem [{\citenamefont {Ikeda}\ \emph {et~al.}(2020)\citenamefont {Ikeda},
  \citenamefont {Matsuda},\ and\ \citenamefont {Sato}}]{IkedaPRL2020}%
  \BibitemOpen
  \bibfield  {author} {\bibinfo {author} {\bibfnamefont {A.}~\bibnamefont
  {Ikeda}}, \bibinfo {author} {\bibfnamefont {Y.~H.}\ \bibnamefont {Matsuda}},\
  and\ \bibinfo {author} {\bibfnamefont {K.}~\bibnamefont {Sato}},\ }\bibfield
  {title} {\bibinfo {title} {\rm{Two spin-state crystallizations in
  LaCoO$_{3}$}},\ }\href {https://doi.org/10.1103/PhysRevLett.125.177202}
  {\bibfield  {journal} {\bibinfo  {journal} {Phys. Rev. Lett.}\ }\textbf
  {\bibinfo {volume} {125}},\ \bibinfo {pages} {177202} (\bibinfo {year}
  {2020})}\BibitemShut {NoStop}%
\bibitem [{\citenamefont {Ikeda}\ \emph {et~al.}(2023)\citenamefont {Ikeda},
  \citenamefont {Matsuda}, \citenamefont {Sato}, \citenamefont {Ishii},
  \citenamefont {Sawabe}, \citenamefont {Nakamura}, \citenamefont {Takeyama},\
  and\ \citenamefont {Nasu}}]{IkedaNC2023}%
  \BibitemOpen
  \bibfield  {author} {\bibinfo {author} {\bibfnamefont {A.}~\bibnamefont
  {Ikeda}}, \bibinfo {author} {\bibfnamefont {Y.~H.}\ \bibnamefont {Matsuda}},
  \bibinfo {author} {\bibfnamefont {K.}~\bibnamefont {Sato}}, \bibinfo {author}
  {\bibfnamefont {Y.}~\bibnamefont {Ishii}}, \bibinfo {author} {\bibfnamefont
  {H.}~\bibnamefont {Sawabe}}, \bibinfo {author} {\bibfnamefont
  {D.}~\bibnamefont {Nakamura}}, \bibinfo {author} {\bibfnamefont
  {S.}~\bibnamefont {Takeyama}},\ and\ \bibinfo {author} {\bibfnamefont
  {J.}~\bibnamefont {Nasu}},\ }\bibfield  {title} {\bibinfo {title}
  {\rm{Signature of spin-triplet exciton condensations in LaCoO$_{3}$ at
  ultrahigh magnetic fields up to 600 T}},\ }\href
  {https://doi.org/10.1038/s41467-023-37125-4} {\bibfield  {journal} {\bibinfo
  {journal} {Nat. Commun.,}\ }\textbf {\bibinfo {volume} {14}},\ \bibinfo
  {pages} {1744} (\bibinfo {year} {2023})}\BibitemShut {NoStop}%
\bibitem [{\citenamefont {Ikeda}\ \emph {et~al.}(2022)\citenamefont {Ikeda},
  \citenamefont {Matsuda}, \citenamefont {Zhou}, \citenamefont {Peng},
  \citenamefont {Ishii}, \citenamefont {Yajima}, \citenamefont {Kubota},
  \citenamefont {Inoue}, \citenamefont {Inubushi}, \citenamefont {Tono},\ and\
  \citenamefont {Yabashi}}]{IkedaAPL2022}%
  \BibitemOpen
  \bibfield  {author} {\bibinfo {author} {\bibfnamefont {A.}~\bibnamefont
  {Ikeda}}, \bibinfo {author} {\bibfnamefont {Y.~H.}\ \bibnamefont {Matsuda}},
  \bibinfo {author} {\bibfnamefont {X.}~\bibnamefont {Zhou}}, \bibinfo {author}
  {\bibfnamefont {S.}~\bibnamefont {Peng}}, \bibinfo {author} {\bibfnamefont
  {Y.}~\bibnamefont {Ishii}}, \bibinfo {author} {\bibfnamefont
  {T.}~\bibnamefont {Yajima}}, \bibinfo {author} {\bibfnamefont
  {Y.}~\bibnamefont {Kubota}}, \bibinfo {author} {\bibfnamefont
  {I.}~\bibnamefont {Inoue}}, \bibinfo {author} {\bibfnamefont
  {Y.}~\bibnamefont {Inubushi}}, \bibinfo {author} {\bibfnamefont
  {K.}~\bibnamefont {Tono}},\ and\ \bibinfo {author} {\bibfnamefont
  {M.}~\bibnamefont {Yabashi}},\ }\bibfield  {title} {\bibinfo {title}
  {\rm{Generating 77 T using a portable pulse magnet for single-shot quantum
  beam experiments}},\ }\href {https://doi.org/10.1063/5.0088134} {\bibfield
  {journal} {\bibinfo  {journal} {Appl. Phys. Lett.}\ }\textbf {\bibinfo
  {volume} {120}},\ \bibinfo {pages} {142403} (\bibinfo {year}
  {2022})}\BibitemShut {NoStop}%
\bibitem [{\citenamefont {Wada}\ \emph {et~al.}(1997)\citenamefont {Wada},
  \citenamefont {Nakamura}, \citenamefont {Mitsuda}, \citenamefont {Shiga},
  \citenamefont {Tanaka}, \citenamefont {Mitamura},\ and\ \citenamefont
  {Goto}}]{WadaJPCM1997}%
  \BibitemOpen
  \bibfield  {author} {\bibinfo {author} {\bibfnamefont {H.}~\bibnamefont
  {Wada}}, \bibinfo {author} {\bibfnamefont {A.}~\bibnamefont {Nakamura}},
  \bibinfo {author} {\bibfnamefont {A.}~\bibnamefont {Mitsuda}}, \bibinfo
  {author} {\bibfnamefont {M.}~\bibnamefont {Shiga}}, \bibinfo {author}
  {\bibfnamefont {T.}~\bibnamefont {Tanaka}}, \bibinfo {author} {\bibfnamefont
  {H.}~\bibnamefont {Mitamura}},\ and\ \bibinfo {author} {\bibfnamefont
  {T.}~\bibnamefont {Goto}},\ }\bibfield  {title} {\bibinfo {title}
  {\rm{Temperature- and field-induced valence transitions of
  EuNi$_2$(Si$_{1-x}$Ge$_x$)$_2$}},\ }\href
  {https://doi.org/10.1088/0953-8984/9/37/021} {\bibfield  {journal} {\bibinfo
  {journal} {J. Phys.: Condens. Matter}\ }\textbf {\bibinfo {volume} {9}},\
  \bibinfo {pages} {7913} (\bibinfo {year} {1997})}\BibitemShut {NoStop}%
\bibitem [{\citenamefont {Matsuda}\ \emph {et~al.}(2009)\citenamefont
  {Matsuda}, \citenamefont {Ouyang}, \citenamefont {Nojiri}, \citenamefont
  {Inami}, \citenamefont {Ohwada}, \citenamefont {Suzuki}, \citenamefont
  {Kawamura}, \citenamefont {Mitsuda},\ and\ \citenamefont
  {Wada}}]{YHMatsudaPRL2009}%
  \BibitemOpen
  \bibfield  {author} {\bibinfo {author} {\bibfnamefont {Y.~H.}\ \bibnamefont
  {Matsuda}}, \bibinfo {author} {\bibfnamefont {Z.~W.}\ \bibnamefont {Ouyang}},
  \bibinfo {author} {\bibfnamefont {H.}~\bibnamefont {Nojiri}}, \bibinfo
  {author} {\bibfnamefont {T.}~\bibnamefont {Inami}}, \bibinfo {author}
  {\bibfnamefont {K.}~\bibnamefont {Ohwada}}, \bibinfo {author} {\bibfnamefont
  {M.}~\bibnamefont {Suzuki}}, \bibinfo {author} {\bibfnamefont
  {N.}~\bibnamefont {Kawamura}}, \bibinfo {author} {\bibfnamefont
  {A.}~\bibnamefont {Mitsuda}},\ and\ \bibinfo {author} {\bibfnamefont
  {H.}~\bibnamefont {Wada}},\ }\bibfield  {title} {\bibinfo {title} {\rm{X-ray
  magnetic circular dichroism of a valence fluctuating state in eu at high
  magnetic fields}},\ }\href {https://doi.org/10.1103/PhysRevLett.103.046402}
  {\bibfield  {journal} {\bibinfo  {journal} {Phys. Rev. Lett.}\ }\textbf
  {\bibinfo {volume} {103}},\ \bibinfo {pages} {046402} (\bibinfo {year}
  {2009})}\BibitemShut {NoStop}%
\bibitem [{\citenamefont {Croft}\ \emph {et~al.}(1982)\citenamefont {Croft},
  \citenamefont {Hodges}, \citenamefont {Kemly}, \citenamefont {Krishnan},
  \citenamefont {Murgai},\ and\ \citenamefont {Gupta}}]{CroftPRL1982}%
  \BibitemOpen
  \bibfield  {author} {\bibinfo {author} {\bibfnamefont {M.}~\bibnamefont
  {Croft}}, \bibinfo {author} {\bibfnamefont {J.~A.}\ \bibnamefont {Hodges}},
  \bibinfo {author} {\bibfnamefont {E.}~\bibnamefont {Kemly}}, \bibinfo
  {author} {\bibfnamefont {A.}~\bibnamefont {Krishnan}}, \bibinfo {author}
  {\bibfnamefont {V.}~\bibnamefont {Murgai}},\ and\ \bibinfo {author}
  {\bibfnamefont {L.~C.}\ \bibnamefont {Gupta}},\ }\bibfield  {title} {\bibinfo
  {title} {\rm{Cooperative Configuration Change in EuPd$_{2}$Si$_{2}$}},\
  }\href {https://doi.org/10.1103/PhysRevLett.48.826} {\bibfield  {journal}
  {\bibinfo  {journal} {Phys. Rev. Lett.}\ }\textbf {\bibinfo {volume} {48}},\
  \bibinfo {pages} {826} (\bibinfo {year} {1982})}\BibitemShut {NoStop}%
\bibitem [{\citenamefont {Raffius}\ \emph {et~al.}(1993)\citenamefont
  {Raffius}, \citenamefont {M\"{o}rsen}, \citenamefont {Mosel}, \citenamefont
  {M\"{u}ler-Warmuth}, \citenamefont {Jeitschko}, \citenamefont
  {Terb\"{u}chte},\ and\ \citenamefont {Vomhof}}]{RaffiusJPCS1993}%
  \BibitemOpen
  \bibfield  {author} {\bibinfo {author} {\bibfnamefont {H.}~\bibnamefont
  {Raffius}}, \bibinfo {author} {\bibfnamefont {E.}~\bibnamefont {M\"{o}rsen}},
  \bibinfo {author} {\bibfnamefont {B.~D.}\ \bibnamefont {Mosel}}, \bibinfo
  {author} {\bibfnamefont {W.}~\bibnamefont {M\"{u}ler-Warmuth}}, \bibinfo
  {author} {\bibfnamefont {W.}~\bibnamefont {Jeitschko}}, \bibinfo {author}
  {\bibfnamefont {L.}~\bibnamefont {Terb\"{u}chte}},\ and\ \bibinfo {author}
  {\bibfnamefont {T.}~\bibnamefont {Vomhof}},\ }\bibfield  {title} {\bibinfo
  {title} {\rm{Magnetic properties of ternary lanthanoid transition metal
  arsenides studied by M\"{o}ssbauer and susceptibility measurements}},\ }\href
  {https://doi.org/10.1016/0022-3697(93)90301-7} {\bibfield  {journal}
  {\bibinfo  {journal} {J. Phys. Chem. Solids}\ }\textbf {\bibinfo {volume}
  {54}},\ \bibinfo {pages} {135} (\bibinfo {year} {1993})}\BibitemShut
  {NoStop}%
\bibitem [{\citenamefont {Raffius}\ \emph {et~al.}(1991)\citenamefont
  {Raffius}, \citenamefont {M\"{o}rsen}, \citenamefont {Mosel}, \citenamefont
  {M\"{u}ller-Warmuth}, \citenamefont {Hilbich}, \citenamefont {Reehuis},
  \citenamefont {Vomhof},\ and\ \citenamefont {Jeitschko}}]{RaffiusJPCS1991}%
  \BibitemOpen
  \bibfield  {author} {\bibinfo {author} {\bibfnamefont {H.}~\bibnamefont
  {Raffius}}, \bibinfo {author} {\bibfnamefont {E.}~\bibnamefont {M\"{o}rsen}},
  \bibinfo {author} {\bibfnamefont {B.~D.}\ \bibnamefont {Mosel}}, \bibinfo
  {author} {\bibfnamefont {W.}~\bibnamefont {M\"{u}ller-Warmuth}}, \bibinfo
  {author} {\bibfnamefont {T.}~\bibnamefont {Hilbich}}, \bibinfo {author}
  {\bibfnamefont {M.}~\bibnamefont {Reehuis}}, \bibinfo {author} {\bibfnamefont
  {T.}~\bibnamefont {Vomhof}},\ and\ \bibinfo {author} {\bibfnamefont
  {W.}~\bibnamefont {Jeitschko}},\ }\bibfield  {title} {\bibinfo {title}
  {\rm{$^{57}$Fe M\"{o}ssbauer spectroscopy and some complementary measurements
  of various ternary iron phosphides}},\ }\href
  {https://doi.org/10.1016/0022-3697(91)90077-d} {\bibfield  {journal}
  {\bibinfo  {journal} {J. Phys. Chem. Solids}\ }\textbf {\bibinfo {volume}
  {52}},\ \bibinfo {pages} {787} (\bibinfo {year} {1991})}\BibitemShut
  {NoStop}%
\bibitem [{\citenamefont {Barla}\ \emph {et~al.}(2004)\citenamefont {Barla},
  \citenamefont {Chefki}, \citenamefont {Huhnt}, \citenamefont {Braden},
  \citenamefont {Leupold}, \citenamefont {R\"{u}ffer}, \citenamefont {Sanchez},
  \citenamefont {Wurth}, \citenamefont {Mewis},\ and\ \citenamefont
  {Abd-Elmeguid}}]{BarlaPRB2004}%
  \BibitemOpen
  \bibfield  {author} {\bibinfo {author} {\bibfnamefont {A.}~\bibnamefont
  {Barla}}, \bibinfo {author} {\bibfnamefont {M.}~\bibnamefont {Chefki}},
  \bibinfo {author} {\bibfnamefont {C.}~\bibnamefont {Huhnt}}, \bibinfo
  {author} {\bibfnamefont {M.}~\bibnamefont {Braden}}, \bibinfo {author}
  {\bibfnamefont {O.}~\bibnamefont {Leupold}}, \bibinfo {author} {\bibfnamefont
  {R.}~\bibnamefont {R\"{u}ffer}}, \bibinfo {author} {\bibfnamefont {J.~P.}\
  \bibnamefont {Sanchez}}, \bibinfo {author} {\bibfnamefont {A.}~\bibnamefont
  {Wurth}}, \bibinfo {author} {\bibfnamefont {A.}~\bibnamefont {Mewis}},\ and\
  \bibinfo {author} {\bibfnamefont {M.~M.}\ \bibnamefont {Abd-Elmeguid}},\
  }\bibfield  {title} {\bibinfo {title} {\rm{Interplay between structural
  instability and lattice dynamics in EuRh$_{2}$(P$_{x}$As$_{1-x}$)$_{2}$}},\
  }\href {https://doi.org/10.1103/PhysRevB.69.100102} {\bibfield  {journal}
  {\bibinfo  {journal} {Phys. Rev. B}\ }\textbf {\bibinfo {volume} {69}},\
  \bibinfo {pages} {100102} (\bibinfo {year} {2004})}\BibitemShut {NoStop}%
\bibitem [{\citenamefont {Momma}\ and\ \citenamefont
  {Izumi}(2011)}]{MommaJAC2011}%
  \BibitemOpen
  \bibfield  {author} {\bibinfo {author} {\bibfnamefont {K.}~\bibnamefont
  {Momma}}\ and\ \bibinfo {author} {\bibfnamefont {F.}~\bibnamefont {Izumi}},\
  }\bibfield  {title} {\bibinfo {title} {\rm{VESTA3 for three-dimensional
  visualization of crystal, volumetric, and morphology data}},\ }\href
  {https://doi.org/10.1107/s0021889811038970} {\bibfield  {journal} {\bibinfo
  {journal} {J. Appl. Crystallogr.}\ }\textbf {\bibinfo {volume} {44}},\
  \bibinfo {pages} {1272} (\bibinfo {year} {2011})}\BibitemShut {NoStop}%
\end{thebibliography}%
\end{document}